# Spatial Control of Charge Doping in n-Type Topological Insulators

*Kazuyuki Sakamoto,*[*,†,‡,¶,§] *Hirotaka Ishikawa,*[¶] *Takashi Wake,*[¶] *Chie Ishimoto,*[¶] *Jun Fujii,*[∥] *Hendrik Bentmann,*[⊥] *Minoru Ohtaka,*[¶] *Kenta Kuroda,*[#] *Natsu Inoue,*[¶] *Takuma Hattori,*[#] *Toshio Miyamachi,*[#] *Fumio Komori,*[#] *Isamu Yamamoto,*[@] *Cheng Fan,*[¶] *Peter Krüger,*[¶,§] *Hiroshi Ota, Fumihiko Matsui, Friedrich Reinert,*[⊥] *José Avila,*[•] *and Maria C. Asensio*[††]

[†]Department of Applied Physics, Osaka University, Osaka 565-0871, Japan

[‡]Center for Spintronics Research Network, Graduate School of Engineering Science, Osaka University, Osaka 560-8531, Japan

[¶]Department of Nanomaterials Science, Chiba University, Chiba 263-8522, Japan

[§]Department of Materials Science and Molecular Chirality Research Center, Chiba University, Chiba 263-8522, Japan

‖Istituto Officina dei Materiali (IOM)-CNR, Laboratorio TASC, I-34149 Trieste, Italy

⊥Experimentelle Physik VII and Röntgen Research Center for Complex Materials, Universität Würzburg, Am Hubland, D-97074 Würzburg, Germany

#Institute for Solid State Physics, The University of Tokyo, Chiba 277-8581, Japan

@Synchrotron Light Application Center, Saga University, Saga 840-8502, Japan

UVSOR Synchrotron Facility, Institute for Molecular Science, Okazaki 444-8585, Japan

•Synchrotron SOLEIL, L'Orme des Merisiers, Saint Aubin-BP 48, 91192 Gif sur Yvette Cedex, France

†† Materials Science Institute of Madrid (ICMM), Spanish Scientific Research Council (CSIC), and the CSIC Associated Unit "MATINÉE," between the Institute of Materials Science of the Valencia University (ICMUV) and the ICMM, E-28049 Cantoblanco, Madrid, Spain

ABSTRACT

Spatially controlling the Fermi level of topological insulators and keeping its electronic states stable are indispensable processes to put this material into practical use for semiconductor spintronics devices. So far, however, such a method has not been established yet. Here we show a novel method for doping hole into n-type topological insulators $Bi_2X_3$ (X= Se, Te) that overcomes the shortcomings of the previous reported methods. The key of this doping is to adsorb $H_2O$ on $Bi_2X_3$ decorated with a small amount of carbon, and its trigger is the irradiation of photon with sufficient energy to excite core-electrons of the outermost layer atoms. This method allows controlling the doping amount by the irradiation time, and acts as photolithography. Such a tunable doping makes it possible to design the electronic states at the nanometer scale, and thus paves a promising avenue toward the realization of novel spintronics devices based on topological insulators.

New states of quantum matter, such as topological insulators (TIs) and Weyl semimetals in which topology, spin-orbit coupling, and broken symmetry play crucial roles,[1] are promising materials to realize future spintronics devices. Among these quantum materials, three-dimensional (3D) TIs are insulating in the bulk but have conducting surface states that are characterized by linearly dispersing helical Dirac fermions with spins locked to momentum and protected from backscattering of non-magnetic impurities by time-reversal symmetry.[1-4] Owing to such properties, 3D TIs provide possible uses for technological applications in spintronics with spin transport without heat dissipation.[5] However, due to naturally generated defects such as vacancies, the bulk of most TIs is carrier doped and therefore conductive.[4] The high bulk conductivity causes surface electron spins to diffuse into the bulk, and thus makes the possibility to fabricate semiconductor spintronics devices based on TI challenging. In order to recover the insulating bulk state, one must compensate the spontaneously generated carriers by adding carriers with opposite sign. In the case of electron doped TIs, divalent metals were used to dope holes into the bulk,[6-11] and gases such as $NO_2$[6,12,13] and $O_2$[9,14,15] were used to hole-dope the surface region of TIs. However, the hole-doped bulk becomes conductive with time due to aging effects,[6] and chemical reaction of gases on the surface is predicted to remove the spin-polarized surface states with Dirac Fermions[16] and moreover it is impossible to control such surface doping under atmospheric air. Consequently, the production of insulating TI bulk, an essential requirement to realize semiconductor spintronics devices, is still an unresolved issue. In this respect, the establishment of a novel method that provides controllable doping and stable doped states is in high demand. Moreover, further establishment of a simple method to control this doping spatially at the nanometer scale will make novel TI based devices more realistic.

Here, we present a complete study on the precise doping control of $Bi_2X_3$ (X= Se and Te), TIs with a single Dirac cone (DC) and large bulk band gap. (The presence of only one DC allows obtaining spin currents with high efficiency by excluding the possibility of inter-DC scattering, and the large band gap is important for the insulation performance.) By combining the adsorption of $H_2O$ and photo-excitation of atoms of the outermost layer, we succeeded to spatially controlling the doping for a sample covered with a certain amount of carbon. $Bi_2X_3$ is formed by X-Bi-X-Bi-X quintuple layers (Figure 1a for $Bi_2Se_3$), and each quintuple layer is weakly bonded by van der Waals interaction, indicating that the outermost layer consists of Se or Te atoms. The present study not only provide an easy and controllable doping method that is stable even in air, but also demonstrate the possibility to create nanometer scale spintronics devices based on topological insulators.

High-quality single crystalline $Bi_2X_3$ samples, grown using the Bridgman method, were cleaved inside UHV chambers using the scotch-tape method under a base pressure of $1 \times 10^{-6}$ Pa to obtain a clean surface before photoelectron spectroscopy (PES) measurement. The band structure of $Bi_2X_3$ was obtained by angle-resolved PES (ARPES) and the core-levels of elements contributing to the doping process were measured by core-level PES (CL-PES). All PES experiments were performed in UHV chambers under a base pressure of $<1 \times 10^{-8}$ Pa at a temperature between 80 K and 100 K. Details of ARPES and CL-PES are described in the Supporting Information.

Figure 1b shows the valence band region of $Bi_2Se_3$ measured with a photon energy ($h\nu$) of 20 eV using ARPES. The bulk conduction band (BCB), clearly observed below the Fermi level ($E_F$), shows that this $Bi_2Se_3$ is initially electron doped, i.e., an n-type TI. The separation in binding energy ($E_B$) between the Dirac point (DP) and the bottom of the BCB (~190 meV) agrees well with the value reported in former ARPES studies.[9-11,14,15,17-23] By continued irradiation with photons of $h\nu$ =20 eV, the DP shifted to lower $E_B$ by approximately 20 meV after 220 min. The shift of DP is more pronounced when using $h\nu$ =100 eV (Figure 1c), where a shift of 185 meV was observed after continuous irradiation for 220 min, the same time as that in Figure 1b, and 240 meV after 490 min. Note that the BCB is also present in the $h\nu$ =100 eV measurement but hardly visible due to matrix element effects. The shift of 240 meV after 490 min indicates that the $E_B$ of the DP ($E_{DP}$) is 145 meV, such that the BCB has moved above the $E_F$, i.e., the bulk band has become insulating. In order to confirm the effect of photon-irradiation on the shift of DP, we performed time-dependent (TD) ARPES measurements at $h\nu =100$ eV by reducing the spot size of the photon beam from 100 μm to 200 nm (Figure 1d). Upon continuous photo-irradiation of 674 min, the DP shifted approximately 210 meV to lower $E_B$ (Figure 1d II). When the position of the light spot was moved by 500 nm (Figure 1d III), i.e., to a fresh area where photons were not irradiated, the DP returned to its original $E_B$ and started to shift to lower $E_B$ again (Figure 1d IV). (Note that almost 100 % of photons are irradiated within a region of 250 nm radius when using a Gaussian photon beam with 200 nm FWHM.) When moving the light spot back to its original position (Figure 1d V), the $E_{DP}$ was the same as before the initial movement. These results indicate that the shift of the DP occurs under photon-irradiation only and that the photo-induced DP shift is a stable and irreversible process. This result contradicts previous studies, in which only a small shift[18] or a DP shift to higher $E_B$ was observed under photo-irradiation[20], and also disagrees with the observation of DP shifting back to higher $E_B$ when stopping the irradiation.[21]

In order to investigate the origin of the photo-induced doping in more detail, and to resolve the discrepancy between the present and former results, we have first performed TD-ARPES measurements using different photon energies. The TD-$E_{DP}$ shows that the DP shift behaves differently depending on the photon energy (Figure 1e). The DP continuously shifts for $h\nu \geq 60$ eV while it saturates at around $E_B$=350 meV for $h\nu \leq 40$ eV. The saturation around $E_B$=350 meV was also reported in former ARPES studies performed with $h\nu = 30$ eV[20] and 23 eV.[21] In Figure 1e, the line curves are least square fits to the experimental data. For the $h\nu \leq$40 eV data, $E_B(t) = A + B\exp(-Ct)$ was used, where A corresponds to the $E_{DP}$ at the saturation condition, B the difference between the initial and saturated $E_{DP}$, and C is a constant corresponding to the mechanism of the TD-$E_{DP}$ shift. We find A=367 meV, B=45 meV, C=0.0187 s$^{-1}$ for $h\nu =20$ eV and A=340 meV, B=108 meV, C=0.0168 s$^{-1}$ for $h\nu$ =40 eV. The almost equal time constant (C) suggests that the hole-doping mechanism is similar at these two photon energies. Taking into account that the DP of a naturally electron doped Bi$_2$Se$_3$ by Se vacancies was mostly observed in a $E_B$ range from 300 to 400 meV, and that the adsorption of residual gases in ultra-high vacuum induces aging effects by doping electrons at the surface region,[6,22,23] we attribute the small difference in A to the different density of Se vacancies and the difference in B to the doping induced by aging. Indeed, the time between the preparation of a fresh surface and the start of the ARPES measurement was longer for $h\nu$ =40 eV than for $h\nu$ =20 eV. In other words, this result suggests that irradiation with $h\nu$ ≤40 eV photon compensates the aging of the sample due to gas adsorption.

In contrast to the case of $h\nu$ ≤40 eV, it is not possible to fit the continuous DP shift at $h\nu$ ≥60 using a single exponential function (Supporting Information, Figure S1). This indicates the presence of more than one doping mechanism. We thus used two exponential functions for the fitting as $E_B(t) = A' + B\exp(-C't) + D'\exp(-E't)$, where D' and E' are parameters for the further hole-doping mechanism. B' varies from 68 to 142 meV depending on the sample similar to B. C', which lies between 0.0148 to 0.0184 s$^{-1}$, agrees with C, meaning that the removal of the aging effect occurs at this photon energy range as well. We found D' = 396±27 meV and E'=1.05 ± 0.13 × 10$^{-3}$ s$^{-1}$. The small value of E' shows that the further hole-doping is a slow process, and the value of A', -100±2 meV, suggests that the DP will move over the Fermi level at the saturation doping. Furthermore, the threshold energy 60 eV, which corresponds to the excitation of the Se 3d core-level, indicates the creation of a core-hole in the outermost layer to be the trigger of this slow hole-doping mechanism.

Further information on the mechanism of photo-induced doping was obtained by CL-PES. Figure 2a shows the Bi 5d, Se 3d, C 1s and O 1s core-levels measured soon after and 400-500 min after starting the irradiation. The four core-levels were measured in the order of Se 3d, Bi 5d, C 1s and O 1s with $h\nu =100$ eV for the former two levels and $h\nu =700$ eV for the latter two. The samples were continuously irradiated during the measurement. The observation of C 1s and O 1s components at the initial stage suggests that the $Bi_2Se_3$ samples used for the results in Figure 1 and 2 were slightly contaminated before the PES measurement. Of the four core-levels, those of Bi, Se and C show TD shifts of approximately 200 meV to lower $E_B$. The approximate parallel $E_B$ shift of the DP and the Se 3d and Bi 5d core-levels clarifies that the present photo-induced doping makes the bulk insulating by bending back the bulk band of the n-type sample as shown in Figure 2b. (In the case of a local hole-doping, the core-level of the positively charged element should shift in the opposite direction, i.e.; to the higher $E_B$ side.) Here, the flat band condition is illustrated by considering the bulk band gap (~300 meV), the gap between the BCB minimum and the DP (~190 meV), and the $E_{DP}$ that is reported to be approximately 100 meV below the $E_F$ from magnetotransport measurements.[24-27] These values indicate the BCB minimum to be located approximately 90 meV above $E_F$, and the bulk valence band (BVB) maximum to be approximately 210 meV below $E_F$ deep in the bulk. The downward band bending of the n-type state, which extends approximately 20 nm into the crystal,[22,23,28] results from electron doping in the surface region that arises from Se vacancies induced when cleaving the sample to obtain a clean surface. (This in-depth length of bulk bending suggests the influence of photo-induced doping to be the 20 nm depth region from the surface, and not the entire depth of the sample.)

By fitting the O 1s core-level spectra, the presence of a large component at $E_B$~533 eV together with two small ones at $E_B$~531 and 529.5 eV is identified at the initial stage. All these three O 1s components shift to lower $E_B$, and the intensity of the two lower $E_B$ components increases drastically by photo-irradiation. Taking the relative $E_B$ and intensity of the three O 1s components into account, and by considering the fact that a component with lower $E_B$ is more negatively charged, the two components with lower $E_B$, or at least one of them is attributed to the donor of hole-doping. In contrast to the previous gas doping studies,[6,8,11,13-15,22,23] no gas was introduced in the vacuum chamber after cleaving the sample in the present study indicating the adsorbed oxygen to come from residual gases. The residual gases that can contribute to the oxidation in a general ultra-high vacuum chamber are $H_2O$, CO and $CO_2$. Exposure to $CO_2$ merely shifts the DP to slightly lower $E_B$, and CO hardly shifts the DP (Supporting Information, Figure S2). This leaves only $H_2O$ as the source of O, which acts as donor. Furthermore, the clear observation of the DC in Figure 1 indicates that the adsorbed O atoms do not suppress the Dirac Fermion as supposed in Reference 16.

Figure 3a and c show the TD-valence bands of two $Bi_2Se_3$ samples measured under a $H_2O$ partial pressure of $1 \times 10^{-6}$ Pa. The sample in Figure 3a was obtained by moving it to the ultrahigh vacuum (UHV) analysis chamber with a base pressure of $1 \times 10^{-8}$ Pa quickly after preparing a clean surface using a Scotch-tape method. The sample in Figure 3c was kept close to the Scotch-tape for a longer time before moving to the analysis chamber like the samples used for the results shown in Figure 1 and 2. As shown in Figure 3b and d, the different preparation leads to a difference in the core-level spectra, i.e., no peak originating from C is observed for the sample in Figure 3b while the C 1s core-level is clearly seen in Figure 3d. This result indicates that tuning the time of keeping the sample close to the scotch tape after the cleaving allows to control the amount of C decorating the surface. Details of the effect of Scotch-tape are given in the Supporting Information (Figure S3). The result in Figure 3a is similar to the former studies in which the DP shifted to higher $E_B$ by $H_2O$ adsorption,[23] but is in contrast to the results shown in Figure 1 and 3c. Taking into account the difference in core-level spectra obtained from the two samples, and the observation of a C 1s peak in the samples used in Figure 1 and 2, we conclude that a small amount of carbon, which was not considered in the former studies, is indispensable for the photo-induced hole-doping mechanism.

Due to the difference in electronegativity of Bi, Se and C,[29-31] the negatively charged surface layers of $Bi_2Se_3$ are expected to turn to a locally positively charged surface by adsorption of carbon, as illustrated in Figure 3e and f. This change in charge state will also modify the adsorption configuration of $H_2O$ molecules. That is, taking the Coulomb interaction into account, the slightly negatively charged O atom of a $H_2O$ molecule will face the C adsorbed surface, while the positively charged H atom will face the surface without C. When a photoelectron is emitted from the outermost Se layer, the Coulomb attraction between the created hole and negatively charged O atom will form a stronger Se-O bonding that is stable even under atmosphere pressure (Supporting Information, Figure S4), while the Coulomb repulsion between the hole and positively charge H atom will release $H_2O$ from the surface.

In conclusion, we have systematically demonstrated that the adsorption of $H_2O$ on a $Bi_2Se_3$ surface decorated with a small amount of carbon is the key for photo-induced hole-doping, and that the excitation of the core-level of the outermost layer atom acts as the trigger. Moreover, the observation of photo-induced hole-doping on $Bi_2Te_3$ (Figure 4a), by irradiating the sample at $hv = 100$ eV, demonstrates this doping mechanism to be a universal mechanism for $Bi_2X_3$. Since the degree of doping can be controlled by the photo-irradiation time for a sample covered with a certain amount of carbon, it is possible to further control the doping spatially by using a kind of photomask like in photolithography. The presented simple recipe for spatially controlled doping of n-type TIs will allow to pattern the wiring for electron spins on TIs (Figure 4b) and/or forming topological p-n junctions[32-36] on TIs at the nanometer scale (Figure 4c), and thus paves a new avenue for the realization of semiconductor spintronics devices using TIs.

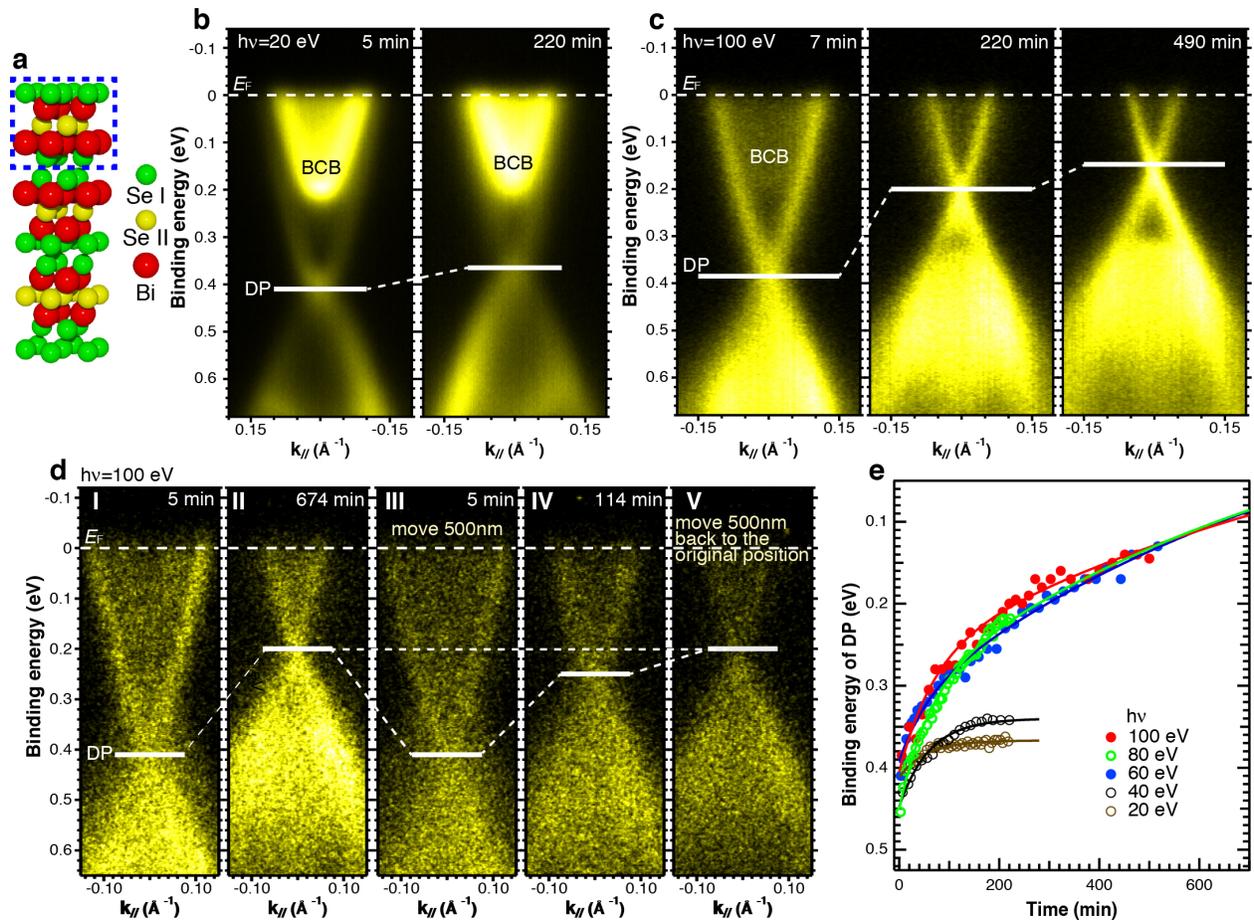

**Figure 1.** (a) Atomic structure of Bi₂Se₃. Red spheres represent the Bi atoms, the green ones are the Se atoms in the outer layers of the quintuple unit, and the yellow spheres are the Se atoms in the middle layer of the unit. The dashed blue box indicates the quintuple unit. TD bulk- and surface-band structure of Bi₂Se₃ measured using light with spot of 100 μm with (b) $h\nu = 20$ eV and (c) $h\nu = 100$ eV. The light was kept irradiating the same sample position with instability of less than 100 nm. The horizontal bars indicate the position of the DP. d) Bulk- and surface-band structure of Bi₂Se₃ measured using light with spot of 200 nm at $h\nu = 100$ eV (panel I). The sample position was moved 500 nm after irradiating 674 min (from panel II to III), and then moved back to the original position after irradiating the new position for 114 min (from panel IV to V). e) TD-$E_{DP}$ measured at $h\nu = 20 - 100$ eV using a light spot of 100 μm. The circles represent the experimental results and the curves overlapping them are the fitting results.

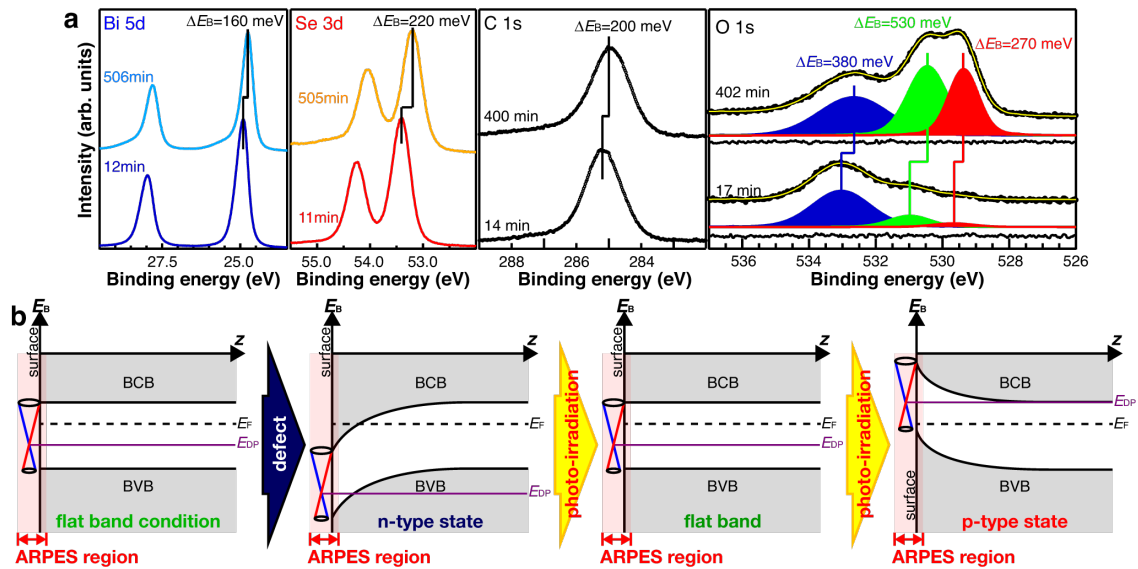

**Figure 2.** (a) Bi 5d and Se 3d core-levels measured using $h\nu = 100$ eV at approximately 10 min after starting the photo-irradiation and approximately 500 min after continuing irradiating the sample, and those of C 1s and O 1s measured at $h\nu = 700$ eV approximately 15 min after starting

the irradiation and approximately 400 min after continuing irradiating the sample. The O 1s core-level spectra were fitted using three components. Dots are the experimental data, the lines overlapping the data are the fitting results and the line under each spectrum is the residue. b) Schematic illustration of the effects of aging and photo-induced doping to the band structure of $Bi_2Se_3$. The $E_{DP}$ is located ~100 meV below the $E_F$ and ~190 meV below the minimum of BCB in the flat band condition. The shaded areas described as "ARPES region" indicate the detectable depth by PES by considering the mean-free path of photoelectron.

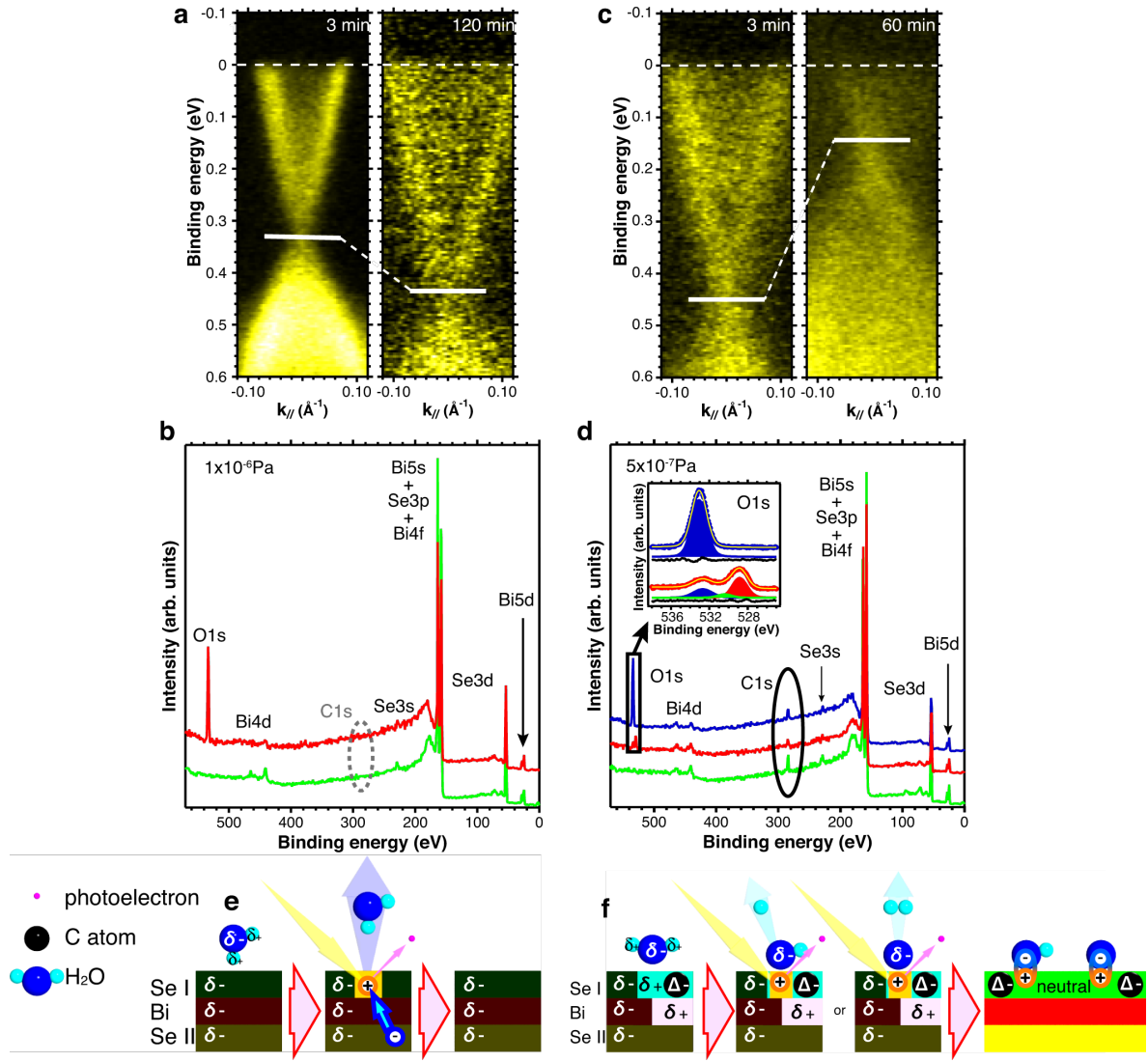

**Figure 3.** (a) TD-ARPES measured at $h\nu =100$ eV under a H$_2$O partial pressure of $1 \times 10^{-6}$ Pa on a Bi$_2$Se$_3$ moved in a UHV chamber with a base pressure of approximately $1 \times 10^{-8}$ Pa quickly after the cleaving and (b) its TD core-level spectra measured at $h\nu =700$ eV. c) The TD-ARPES and (d) TD core-level spectra of Bi$_2$Se$_3$ kept close to the scotch-tape for a longer time after the cleaving and measured with the same condition as those of (a) and (b). The green curves in (b) and (d) have been measured approximately 10 min after starting the irradiation, and the red curve have been measured 130 min and 60 min after continuing irradiating the sample at $h\nu =100$ eV in (b) and (d), respectively. The blue curve in (d) was obtained after moving the light spot 2 mm away from the original position. The elliptical dashed line in (c) suggests the place where the C 1s core-level should be observed, and the elliptical solid line and the rectangle in (d) indicate the C 1s and O 1s core-levels, respectively. Schematic illustration of photo-induced reaction of physisorbed H$_2$O in case of (e) carbon free Bi$_2$Se$_3$ and (f) Bi$_2$Se$_3$ slightly decorated with carbon.

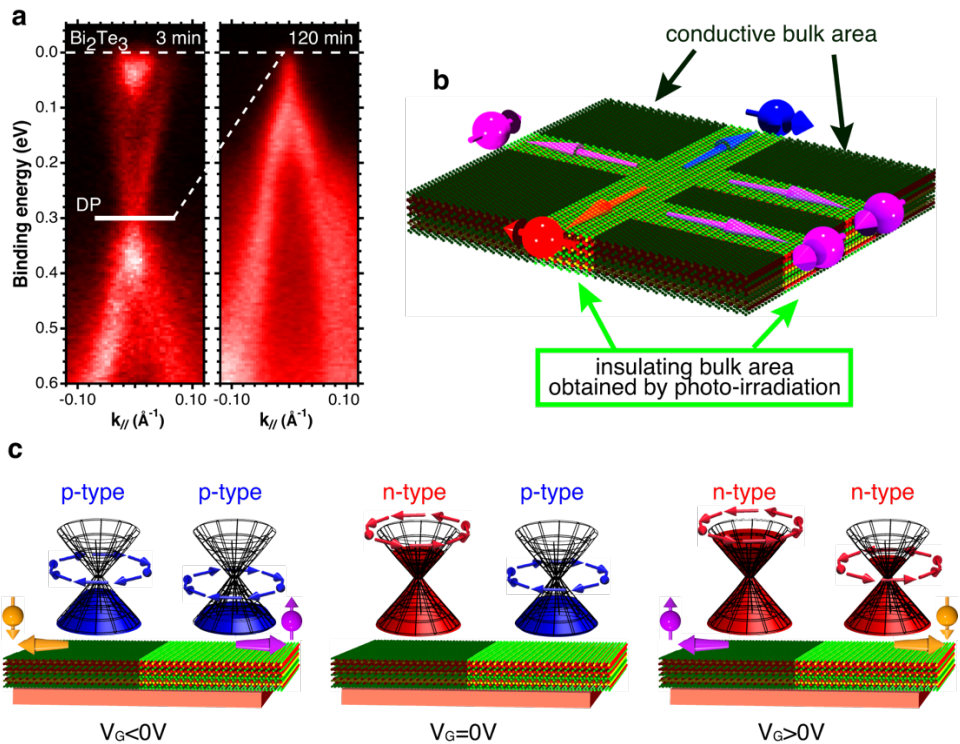

**Figure 4.** (a) TD-ARPES of a $Bi_2Te_3$ sample slightly decorated with C atom. The measurements were performed using $hv = 100$ eV. Illustrations of (b) a circuit for semiconductor spintronics devices and (c) a spin transistor based on topological p-n junction obtained by patterning TI using the photo-induced hole-doping method.

ASSOCIATED CONTENT

**Supporting Information**.

The Supporting Information is available free of charge at

https://pubs.acs.org/doi/10.1021/acsnano.XXXXXXX.

Details of ARPES and CL-PES experiments, fitting of TD-$E_{DP}$ of $Bi_2Se_3$ measured at $hv = 60$ eV, the effect of CO and $CO_2$, the origin of the carbon decorating the $Bi_2Se_3$ sample, and the spatially-resolved mapping of O on photo-induced hole-doped $Bi_2Se_2$. (PDF)

AUTHOR INFORMATION

**Corresponding Author**

Kazuyuki Sakamoto - Department of Applied Physics, Osaka University, Osaka 565-0871, Japan; Center for Spintronics Research Network, Graduate School of Engineering Science, Osaka University, Osaka 560-8531, Japan; Department of Nanomaterials Science, Chiba University, Chiba 263-8522, Japan; Department of Materials Science and Molecular Chirality Research Center, Chiba University, Chiba 263-8522, Japan; orcid.org/0000-0001-9507-6435; E-mail: kazuyuki_sakamoto@ap.eng.osaka-u.ac.jp

**Authors**

Hirotaka Ishikawa – Department of Nanomaterials Science, Chiba University, Chiba 263-8522, Japan

Takashi Wake – Department of Nanomaterials Science, Chiba University, Chiba 263-8522, Japan

Chie Ishimoto – Department of Nanomaterials Science, Chiba University, Chiba 263-8522, Japan

Jun Fujii – Istituto Officina dei Materiali (IOM)-CNR, Laboratorio TASC, I-34149 Trieste, Italy; orcid.org/0000-0003-3208-802X

Hendrik Bentmann – Experimentelle Physik VII and Röntgen Research Center for Complex Materials, Universität Würzburg, Am Hubland, D-97074 Würzburg, Germany

Minoru Ohtaka – Department of Nanomaterials Science, Chiba University, Chiba 263-8522, Japan

Kenta Kuroda – Institute for Solid State Physics, The University of Tokyo, Chiba 277-8581, Japan; orcid.org/0000-0002-0151-0876

Natsu Inoue – Department of Nanomaterials Science, Chiba University, Chiba 263-8522, Japan

Takuma Hattori – Institute for Solid State Physics, The University of Tokyo, Chiba 277-8581, Japan; orcid.org/0000-0003-0339-1847

Toshio Miyamachi – Institute for Solid State Physics, The University of Tokyo, Chiba 277-8581, Japan; orcid.org/0000-0002-9623-1190

Fumio Komori – Institute for Solid State Physics, The University of Tokyo, Chiba 277-8581, Japan; orcid.org/0000-0002-6405-4177

Isamu Yamamoto – Synchrotron Light Application Center, Saga University, Saga 840-8502, Japan

Cheng Fan – Department of Nanomaterials Science, Chiba University, Chiba 263-8522, Japan

Peter Krüger – Department of Nanomaterials Science, Chiba University, Chiba 263-8522, Japan Department of Materials Science and Molecular Chirality Research Center, Chiba University, Chiba 263-8522, Japan; orcid.org/0000-0002-1247-9886

Hiroshi Ota – UVSOR Synchrotron Facility, Institute for Molecular Science, Okazaki 444-8585, Japan

Fumihiko Matsui – UVSOR Synchrotron Facility, Institute for Molecular Science, Okazaki 444-8585, Japan; orcid.org/0000-0002-0681-4270

Friedrich Reinert – Experimentelle Physik VII and Röntgen Research Center for Complex Materials, Universität Würzburg, Am Hubland, D-97074 Würzburg, Germany

José Avila – Synchrotron SOLEIL, L'Orme des Merisiers, Saint Aubin-BP 48, 91192 Gif sur Yvette Cedex, France; orcid.org/0000-0003-1027-5676

Maria C. Asensio – Materials Science Institute of Madrid (ICMM), Spanish Scientific Research Council (CSIC), and the CSIC Associated Unit "MATINÉE," between the Institute of Materials Science of the Valencia University (ICMUV) and the ICMM, E-28049 Cantoblanco, Madrid, Spain; orcid.org/0000-0001-8252-7655

**Author Contributions**

K.S.; H.I.; J.F.; H.B.; M.O.; J.A.; and M.C.A. carried out the nano-TD-ARPES and CL-PES measurements at SOLEIL. K.S.; T.W.; C.I. and I.Y. performed PES measurements under high $H_2O$ partial pressure at SAGA Light Source, and K.S.; H.I and J.F. carried out TD-ARPES and CL-PES measurements at Elettra, MAX-lab and KEK-PF. K.K. created the sample. K.S.; T.W.; H.O. and F.M. did spatial-resolved PES at SPring-8. N.I; T.H.; T.M. and F.K. carried out the mass spectroscopy measurement and scanning tunneling measurement. C.F. and P.K. performed the STM simulations. K.S. analyzed the data and wrote the manuscript with input from H.I.; T.W.; J.F.; H.B.; K.K.; T.M.; F.K.; I.Y.; P.K.; F.M.; F.R and M.C.A. K.S. conceived and coordinated the project. All authors discussed the results and commented on the manuscript.

**Notes**

The authors declare no competing financial interest.

ACKNOWLEDGMENT

This research is supported by the JSPS Grant-in-Aid for Scientific Research (B) JP25287070 and JP19H02592, the JSPS Grant-in-Aid for Specially Promoted Research JP20H05621, the JSPS Grant-in-Aid for Scientific Research on Innovative Areas "3D Active-Site Science" JP26105003 and JP17H05211, and was performed with the approval of the Japan Synchrotron Radiation Research Institute (Proposal No.2017B0124 and 2017B1518). This work is also partly funded by the Deutsche Forschungsgemeinschaft (DFG, German Research Foundation) through project - ID 258499086 - SFB 1170 (A01), the Würzburg-Dresden Cluster of Excellence on Complexity and Topology in Quantum Matter – ct.qmat Project-ID 390858490 - EXC 2147.